\documentclass[aps,prb,twocolumn,superscriptaddress,floatfix]{revtex4}

\usepackage{graphicx}

\begin{document}
\title{Field Induced Spin Supersolidity in Frustrated Spin-1/2 Spin-Dimer Models}

\author{Pochung Chen}
\email{pcchen@phys.nthu.edu.tw} %
\affiliation{Department of Physics, National Tsing Hua University,
Hsinchu 30013, Taiwan}

\author{Chen-Yen Lai} %
\affiliation{Department of Physics, National Tsing Hua University,
Hsinchu 30013, Taiwan} %
\affiliation{Department of Physics and Astronomy, University of
California, Riverside, California 92521, USA}

\author{Min-Fong Yang}
\affiliation{Department of Physics, Tunghai University, Taichung
40704, Taiwan}

\date{\today}

\begin{abstract}
By means of the recently developed algorithm based on the tensor
product states, the magnetization process of frustrated spin-$1/2$
spin-dimer models on a square lattice is investigated. Clear
evidence of a supersolid phase over a finite regime of magnetic
field is observed. Besides, critical fields at various
field-induced transitions are determined accurately. Our work
hence sheds new light on the search of the supersolid phase in
real frustrated spin-dimer compounds.

\end{abstract}


\maketitle

Over the last decade, dimerized quantum antiferromagnets have
attracted much attention, since rich field-induced quantum phases
can appear in these compounds.~\cite{Giamarchi08} In the absence
of magnetic field, the system exhibits spin-singlet ground state,
consisting of dimers for closely spaced pairs of spins $S=1/2$. By
applying a large enough field such that the energy gap to
spin-triplet excitations closes, the lowest triplet excitation
starts to condense and the system develops a spin superfluid (SF)
state, which supports a staggered magnetization transverse to the
field direction. When the effective repulsion between triplets
overwhelms their kinetic energy, instead of Bose-Einstein
condensation of triplet excitations, incompressible commensurate
crystals of triplets with broken translational symmetry can be
stabilized,~\cite{Rice02} which are signaled by magnetization
plateaus.

Recently, an even more exotic {\it supersolid} (SS) phase is found
in the vicinity of a magnetization plateau for some spin-dimer
models.~\cite{Ng06} This generated an enormous interest in the
study of quantum and thermal phase transitions out of these spin
SS states.~\cite{Laflorencie07,spin_dimer_2,Sengupta07} The main
character of the SS phase is that its ground states possess both
solid and SF long range orders. It has been proposed that
correlated hopping of triplets may play a crucial role in forming
this novel phase.~\cite{Schmidt08} At the SF-SS transition point,
a magnetization anomaly and thereby a discontinuity in magnetic
susceptibility appears. This indicates that the SF-SS transition
is of second order. Moreover, approaching the edge of the
magnetization plateau from the SS phase, the spin stiffness
vanishes linearly,~\cite{Laflorencie07} in agreement with the
superfluid-insulator universality class.~\cite{SF-MI}

For the spin-$1/2$ spin-dimer models studied previously, large
Ising-like exchange anisotropy is necessary for the existence of
the SS phase,~\cite{Ng06,Laflorencie07,spin_dimer_2} which is
unrealistic for most magnetic systems. Under some kind of
approximation, such an anisotropy can be considered as an
effective interaction resulting from frustrated spin-isotropic
couplings.~\cite{Ng06,Sengupta07} Thus the results for the
spin-anisotropic cases suggest that the SS phase can appear also
in frustrated spin-isotropic models. Nevertheless, because of
underlying approximation, quantitative predictions of the needed
frustrated coupling and magnetic field for the appearance of this
novel phase can not be determined accurately in the preceding
studies. To provide useful guide to the experimental search of the
SS phase in real frustrated spin-dimer compounds, investigations
directly within the frustrated spin-isotropic models are called
for.

In this paper, various field-induced quantum phase transitions in
frustrated spin-$1/2$ spin-dimer Heisenberg models are explored by
using the combined algorithm~\cite{Jiang08} of the infinite
time-evolving block decimation (iTEBD) method~\cite{iTEBD} and the
tensor renormalization group (TRG) approach.~\cite{TRG} In this
combined algorithm, the ground states are assumed in the form of
the tensor product state (TPS) or the projected entangled-pair
state (PEPS).~\cite{TPS_review}
The power of the TPS/PEPS-based approach in studying
first-order quantum phase transitions has been demonstrated by
several groups.~\cite{Orus09,Chen09,Bauer09} We note that, in
contrast to quantum Monte Carlo simulation which is plagued by the
sign problem in studying frustrated spin systems, the numerical
approach employed here is appropriate
because frustration does not introduce additional difficulties to the
TPS/PEPS-based method.~\cite{Bauer09,PEPS-frustrated} By using the
combined iTEBD and TRG method,
the existence of the SS phase in the frustrated systems under
consideration is firmly established, and the critical fields
bounding this phase are determined precisely. These results
provide further theoretical support on searching the SS phases in
the neighborhood of magnetization plateaus observed in some spin
dimer compounds with frustration. Moreover, the success in
obtaining precise results for frustrated spin systems clearly
demonstrates that the combined algorithm can be an efficient and
accurate numerical tool with wide applications including even
frustrated systems.

\begin{figure}[tb]
\includegraphics[width=1.54in]{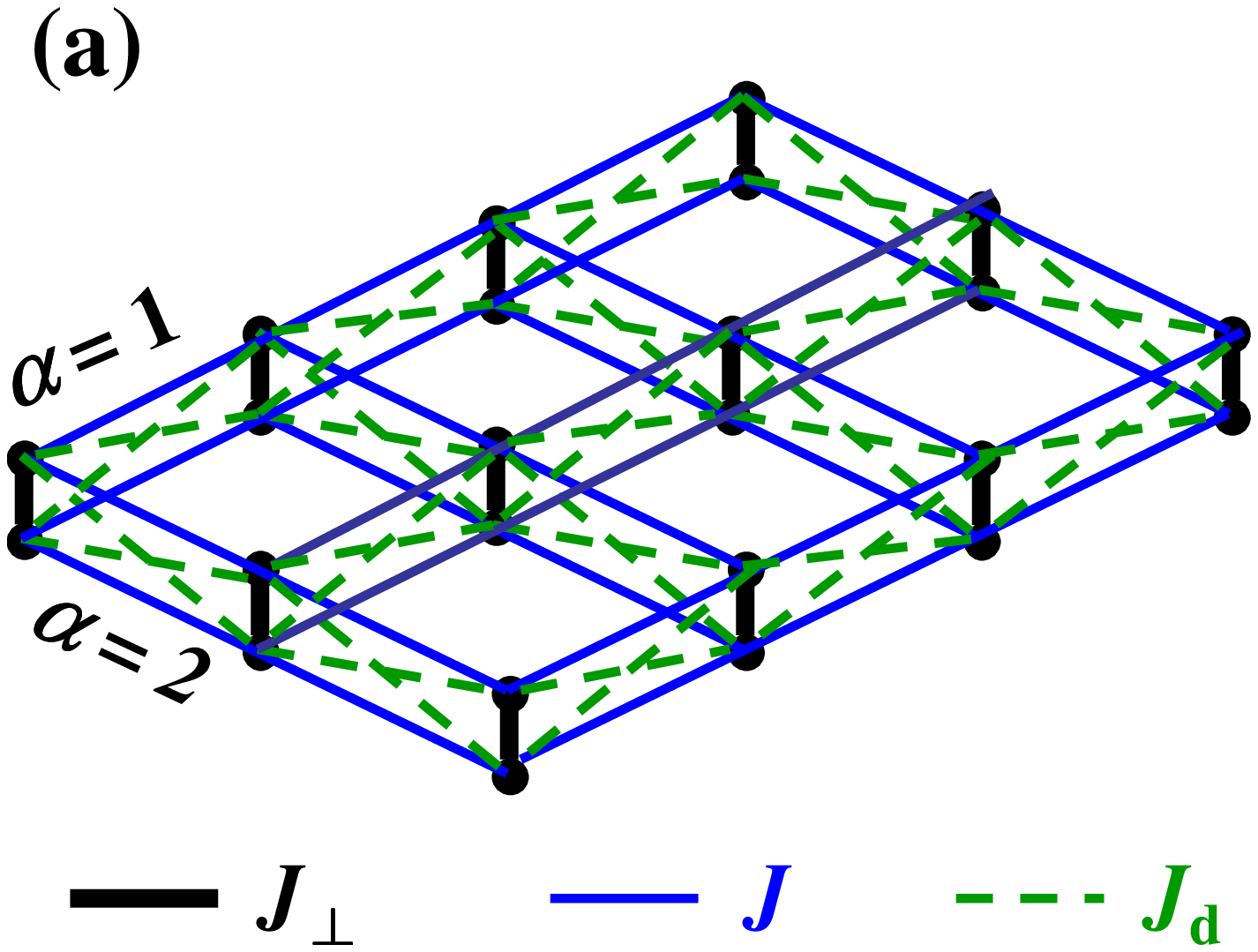} \hspace{0.6cm}
\includegraphics[width=1.54in]{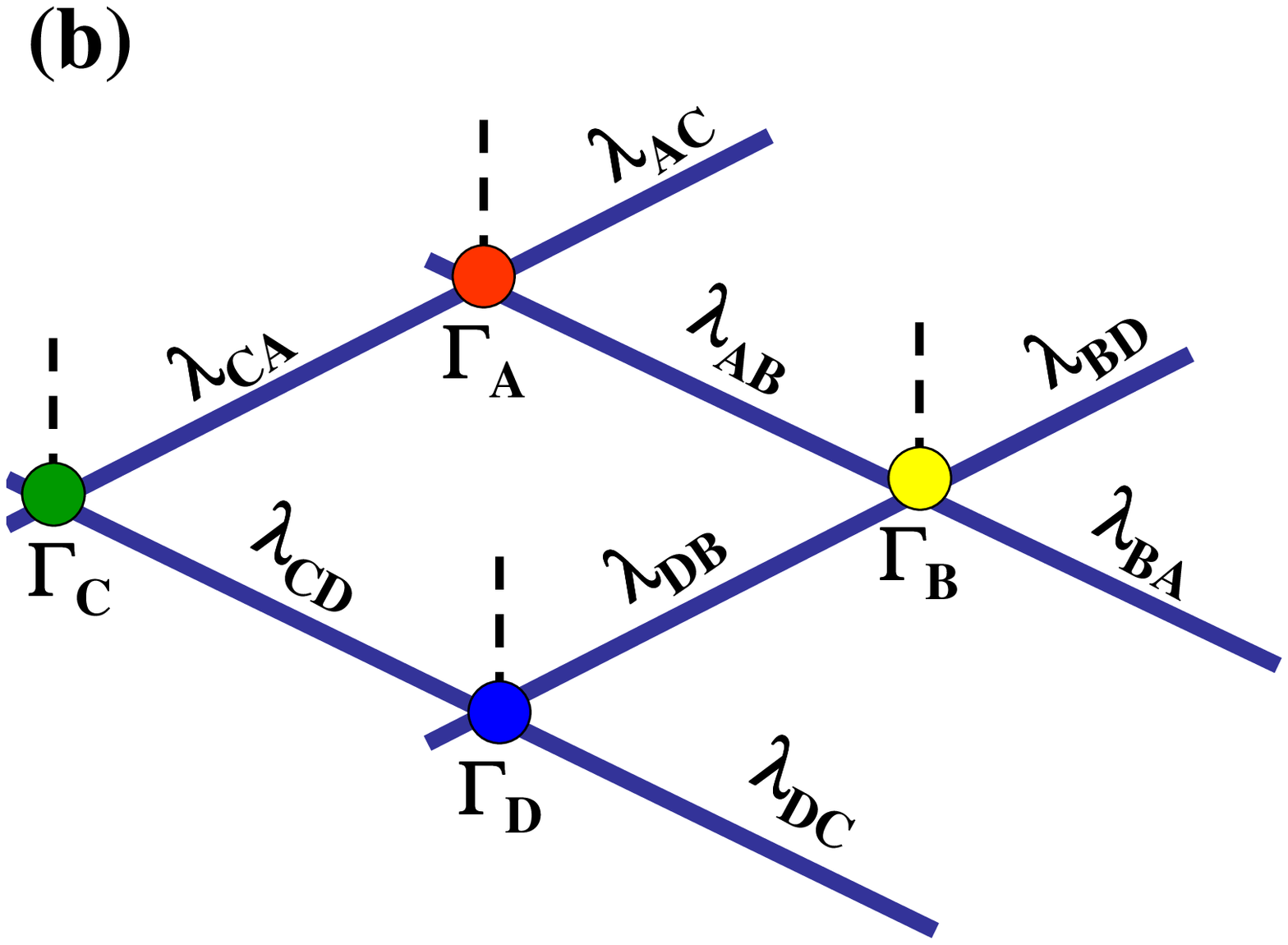}
\caption{(Color online) (a) Square lattice of $S=1/2$ dimers with
an intra-dimer interaction $J_\bot$ and inter-dimer interactions
$J$ and $J_d$. (b) The construction of periodic tensor network
with a $2\times2$ unit cell.} \label{fig:model}
\end{figure}

We consider the following spin-$1/2$ bilayer Heisenberg
Hamiltonian under a uniform magnetic field $h$ on a square lattice
with a strong interlayer exchange $J_\bot$, a weaker intralayer
coupling $J$, and a frustrating interlayer interaction $J_d$, [see
Fig.~\ref{fig:model}(a)]
\begin{eqnarray}
H &=& J_\bot \sum_{i} {\bf S}_{i,1} \cdot {\bf S}_{i,2} - h
\sum_{i,\alpha} S^z_{i,\alpha}  \nonumber \\
&& + J \sum_{\langle i,j \rangle, \alpha} {\bf S}_{i,\alpha} \cdot
{\bf S}_{j,\alpha} + J_d \sum_{\langle i,j \rangle, \alpha} {\bf
S}_{i,\alpha} \cdot {\bf S}_{j,\bar{\alpha}} \; . \label{eq:Hami}
\end{eqnarray}
Here the antiferromagnetic couplings are assumed ($J_\bot$, $J$,
$J_d > 0$). The indices $i$ and $j$ denote rungs of the bilayer
lattice, and $\alpha=1$, 2 labels the two different layers. The
summation $\langle i,j \rangle$ runs over pairs of
nearest-neighbor rungs. Henceforth, $J_\bot \equiv 1$ is set to be
the energy unit. To characterize different phases, several local
order parameters are calculated.
Firstly, when systems undergo the dimer-SF transition, the
$z$-component uniform magnetization per site $m^z_{\rm u} \equiv
(1/2N) \sum_i \left\langle S^z_{i,1} + S^z_{i,2} \right\rangle$
begins to be nonzero, where $N$ is the total number of rungs on
the bilayer.
Secondly, the SF states with nonzero in-plane antiferromagnetic
magnetization along the $x$ direction can be detected by the
$x$-component staggered magnetization per site $m^x_{\rm st}
\equiv (1/2N) \sum_i \left\langle S^x_{i,1} - S^x_{i,2}
\right\rangle e^{i{\bf Q}\cdot{\bf r_i}}$. Using the mapping for
the triplets to the semihardcore bosons $b^\dag_i = \left(
S^+_{i,1} - S^+_{i,2} \right) e^{i{\bf Q}\cdot{\bf r_i}}
/\sqrt{2}$,~\cite{Ng06} the condensate density $n_0$ of triplet
excitations can be related to $m^x_{\rm st}$ by $n_0 \equiv
|\langle b^\dag \rangle|^2 = 2|m^{x}_{\rm st}|^2$.
Finally, the checkerboard solid (CBS) order will be signaled by a
finite value of the $z$-component staggered magnetization per site
$m^{z}_{\rm st} \equiv (1/2N) \sum_i \left\langle S^z_{i,1} +
S^z_{i,2} \right\rangle e^{i{\bf Q}\cdot{\bf r_i}}$, where ${\bf
Q}=(\pi,\pi)$. In the thermodynamic limit, this quantity has a
direct relation, $|m^{z}_{\rm st}|^2 =S({\bf Q})/2N$, to the
static structure factor $S({\bf Q}) \equiv (1/2N)\sum_{i,j}
\left\langle (S^z_{i,1} + S^z_{i,2}) (S^z_{j,1} + S^z_{j,2})
\right\rangle e^{i{\bf Q}\cdot({\bf r_i}-{\bf r_j})}$.

These quantities are evaluated under the combined iTEBD and TRG
algorithm,~\cite{Jiang08,Chen09} where the ground-state wave
function is approximated by a TPS/PEPS ansatz.~\cite{TPS_review}
Our construction of TPS for these bilayer systems on a
square lattice is to attach a rank-five tensor
$[\Gamma_i]^s_{lrud}$ to each {\it rung} $i$ and a diagonal
singular value matrix 
$[\lambda_{\langle i,j\rangle}]_l$ to each bond of
nearest-neighboring rungs $i$ and $j$, as sketched in
Fig.~\ref{fig:model}(b). Here $s$ is the physical index with
$s=1,\cdots,4$ for the present spin-dimer case, and $l,r,u,d(=1
\cdots D)$ denote the virtual bond indices in four directions. The
approximation can be systematically improved simply by increasing
the bond dimension $D$ of the underlying tensors. The optimized
TPS/PEPS is determined through the power method via iterative
projections for a given initial state. This procedure can be
considered as a generalization of the one-dimensional iTEBD
method~\cite{iTEBD} to the two dimensional cases. To improves the
stability the algorithm, the calculated ground state for a lower
field $h$ is usually employed as the initial state for the
higher-$h$ case.~\cite{note} By means of TRG method,~\cite{TRG}
the expectation values for a TPS/PEPS ground state of very large
systems can be calculated efficiently, where the accuracy can be
systematically improved by increasing the TRG cutoff $D_{\rm
cut}$. In the present work, we consider the bond dimension up to
$D=5$ and keep $D_{\rm cut}\ge D^2$ to ensure the accuracy of the
TRG calculation.

\begin{figure}[tb]
\includegraphics[width=3in]{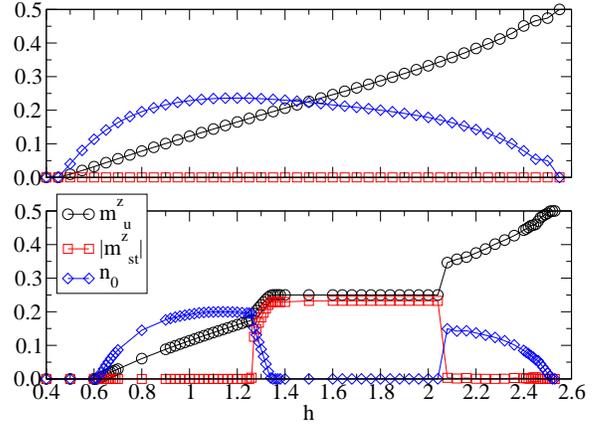}
\caption{(Color online) Values of $m^z_{\rm u}$, $|m^{z}_{\rm
st}|$, and $n_0=2|m^x_{\rm st}|^2$ for the ground states at
$J_d=0.15$ (upper panel) and 0.21 (lower panel) with $J = 0.38$ as
functions of external field $h$ for systems of size $2^7\times
2^7$ with $D=4$ and $D_{\rm cut}=16$. The $J_d$ independence of
the saturation field $h_s$ can be observed. } \label{fig:full}
\end{figure}


The results of the local order parameters defined above as
functions of magnetic field $h$ for $J_d = 0.15$ and 0.21 with $J
= 0.38$ and systems size $N=2^7\times 2^7$ are shown in
Fig.~\ref{fig:full}. Here we take the bond dimension $D=4$ and the
TRG cutoff $D_{\rm cut}=16$.
We note that results for $D=3$, 5 are very similar to those for $D=4$ (see below).
Upon increasing the field, the ground state of the model in
Eq.~(\ref{eq:Hami}) goes through a succession of distinct quantum
phases. For the case of weak frustration ($J_d= 0.15$), the ground
state evolves first from the dimer phase ($m^z_{\rm u}$, $n_0=0$)
to the SF phase ($m^z_{\rm u}$, $n_0\neq 0$), and finally to the
fully-polarized (FP) phase ($m^z_{\rm u}=1/2$ and $n_0=0$). Such a
behavior is consistent with the results of the unfrustrated ($J_d
= 0$) case.~\cite{unfrustrated}

For the larger-$J_d$ case ($J_d= 0.21$), apart from aforementioned
phases, a magnetization plateau and a peculiar SS phase with both
the CBS and the SF orders are found. The general feature of our
results for $J_d= 0.21$ is quite similar to what have been
discovered in the previous works for the spin anisotropic
systems.~\cite{Ng06,Laflorencie07,spin_dimer_2} Within the regime
of magnetization plateau, we find that the uniform magnetization
is $m^z_{\rm u} \simeq 0.25$ together with a nonzero CBS order
parameter $|m^{z}_{\rm st}| \simeq 0.23$. Thus this plateau state
describes the CBS phase with long-range diagonal order at wave
vector ${\bf Q}=(\pi,\pi)$. The observed CBS phase can be
understood as formed by half-filling the system with triplet
excitations so that $m^z_{\rm u} = 0.25$. Its stability comes from
strong effective repulsion between nearest neighbor triplets
generated by interdimer frustrated couplings $J$ and $J_d$. We
note that, as in the anisotropic cases studied
previously,~\cite{Ng06,Laflorencie07,spin_dimer_2} the calculated
CBS order parameter $|m^{z}_{\rm st}|$ is also slightly smaller
than its classical value $|m^{z}_{\rm st}|= 1/4$, which is caused
by quantum fluctuations therein. At higher fields, the melting of
the CBS is found to be of first order and the SF phase reappears.
At this transition, the magnetization changes abruptly from
$m^z_{\rm u} \simeq 0.25$ to $m^z_{\rm u} \simeq 0.35$, and the
condensate fraction $n_0$ jumps from zero to $n_0 \simeq 0.15$. On
the other hand, when decreasing the field from the plateau, a SS
phase emerges in the present frustrated spin-dimer system, where
both the CBS order parameter $m^{z}_{\rm st}$ {\it and} the Bose
condensate density $n_0$ are nonzero. At even lower fields, the
CBS order disappears and a standard SF phase with only $n_0\neq 0$
is recovered.

\begin{figure}[tb]
\includegraphics[width=3in]{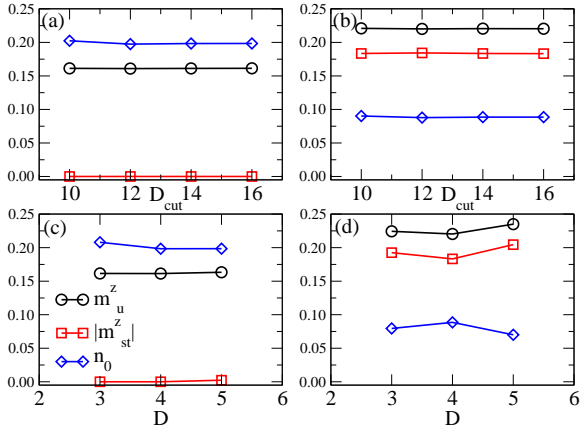}
\caption{(Color online) Dependence of the calculated order
parameters on $D_{\rm cut}$ (upper panels) and $D$ (lower panels):
(a) $h=1.2$ and (b) 1.3 with $D=4$; (c) $h=1.2$ and (d) 1.3 with
$D_{\rm cut}=16$ for $D\le 4$ and $D_{\rm cut}=25$ for $D=5$.
\label{fig:convergence} }
\end{figure}

As a check on our prescription of TRG cutoff $D_{\rm cut}$ and
bond dimension $D$, the dependence of the evaluated order
parameters on them is shown in Fig.~\ref{fig:convergence}. Here we
consider two typical values of $h=1.2$ and 1.3, which are
corresponding to the SF and the SS phases respectively. It is
expected that the ground states in these two cases are more
entangled than those in the dimer or the CBS phases. Consequently
reliable results can be reached only when $D$ and $D_{\rm cut}$
are large enough. As seen from Fig.~\ref{fig:convergence}, there
is almost no dependence on $D_{\rm cut}$ from $D_{\rm cut}=10$ to
$16$ for $D=4$, and the dependence on $D$ is small. (The variance
of each order parameter is less than 0.025 even in the SS phase
with $h = 1.3$.) These results indicate that the bond dimension
and TRG cutoff used in this work is large enough to provide
accurate findings for the present case.

\begin{figure}[tb]
\includegraphics[width=3in]{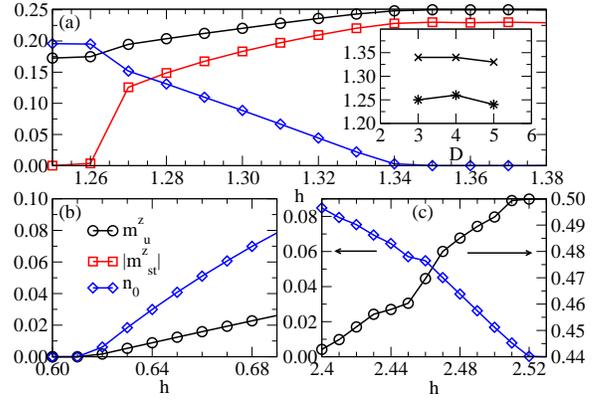}
\caption{(Color online) Details of local order parameters
around (a) the SS phase, (b) the dimer-SF and (c) the SF-FP
transition points with enlarged scale. The parameters are the same
as those for the lower panel of Fig.~\ref{fig:full}. The inset in
panel (a) shows the critical fields $h_{c2}$ ($\times$) and
$h_{c3}$ ($*$) for various $D$  ($D_{\rm cut}=16$ for $D\le 4$ and
$D_{\rm cut}=25$ for $D=5$).
\label{fig:enlarged}}
\end{figure}

To explore the behaviors around the transition points to the SS
phase, results around this phase with enlarged scale are shown in
Fig.~\ref{fig:enlarged}(a). We find that the transition between
the SF and the SS phases occurs at the critical field
$h_{c2} \approx 1.26$, 
at which the CBS order $m^{z}_{\rm st}$ begins to vanish. As
observed in Fig.~\ref{fig:enlarged}(a) (also in
Fig.~\ref{fig:full}), this transition is also signaled by a kink
in the magnetization curve, which implies a discontinuity in
magnetic susceptibility $\chi\equiv d m^z_{\rm u}/dh$. The
field-induced transition between the SS and the CBS phases occurs
at another critical field
$h_{c3} \approx 1.34$, 
at which the condensate fraction $n_0$ begins to vanish. Moreover,
we observe that $n_0$ vanishes linearly as $h$ approaches to
$h_{c3}$ from below. This behavior is in agreement with the
SF-insulator universality class,~\cite{SF-MI} since the density of
the bosonic excitations (triplet {\it holes}) should be small
around $h_{c3}$ and the condensate fraction roughly equals to the
spin stiffness in this dilute-boson limit. In the inset of
Fig.~\ref{fig:enlarged}(a), the dependence of the critical fields
$h_{c2}$ and $h_{c3}$ on bond dimension $D$ is presented. The
values show only a minor dependence on $D$,
which again supports the validity of the present approach.

Now we turn our attention to the cases near the dimer-SF
transition at the critical field $h_{c1}$ and around the
saturation field $h_{s}$. In both cases, some analytical results
are available. As same as the unfrustrated ($J_d = 0$)
case,~\cite{unfrustrated} the instability of the dimer phase is
triggered by energy gap closing of the bosonic {\it one-triplet}
excitation above the dimer state. On the other hand, the FP state
becomes unstable when the lowest bosonic excited state {\it with a
single spin flip} is degenerate with the FP state. Thus both
transitions are expected to be of second order and in the
universality class of the dilute Bose gas quantum-critical
point.~\cite{Sachdev:book} By using the expression of the
excitation spectrum of the one-triplet states up to the
third-order perturbation expansion,~\cite{Gu00} the critical field
$h_{c1}$ of the dimer-SF transition is achieved, %
$h_{c1} \simeq 1 - 2(J-J_d) - \frac{3}{2}J(J-J_d)^2$. To determine
the saturation field $h_{s}$, one notes that the states with a
single spin flip are the eigenstates of our model in
Eq.~(\ref{eq:Hami}) and their energy eigenvalues can be calculated
{\it exactly}. Thus the exact expression for $h_{s}$ can be
derived. We find that, when $J_d < J$ and $J_d < 1/4$, %
the saturation field $h_s = 1 + 4J$, which is independent of the
magnitude of the frustrated coupling $J_d$. A closer look around
the transitions at $h_{c1}$ and $h_{s}$ is shown in
Figs.~\ref{fig:enlarged}(b) and (c). In agreement with the
SF-insulator universality class,~\cite{SF-MI} we find that the
condensate fraction $n_0$ vanishes linearly also around these two
field-induced transition points. The dimer-SF transition occurs at
$h_{c1} \approx 0.61$, where the uniform magnetization $m^z_{\rm
u}$ and the condensate fraction $n_0$ start to be nonzero. Above
the saturation field $h_{s} \approx 2.52$, all spins are polarized
in the $z$ direction and the uniform magnetization saturates at
the value of $m^z_{\rm u}=0.5$. Our findings of $h_{c1}$ and
$h_{s}$ agree well with the values ($h_{c1} \simeq 0.64$ and
$h_{s} = 2.52$) given by the analytic formulas with $J = 0.38$ and
$J_d = 0.21$. Moreover, as seen from Fig.~\ref{fig:full}, the
calculated values of the saturation field $h_{s}$ for the cases of
$J_d = 0.15$ and 0.21 are identical. Thus the independence of the
saturation field on the frustrated coupling $J_d$ is indeed
preserved in our approach. These observations further substantiate
that the employed method can provide reliable results even for the
frustrated spin systems.


\begin{figure}[tb]
\includegraphics[width=3in]{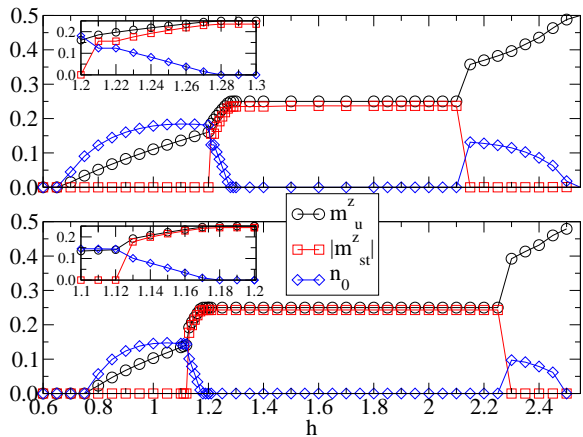}
\caption{(Color online) Local order parameters at $J_d = 0.23$
(upper panel) and 0.27 (lower panel) as functions of external
field $h$. Other parameters are the same as those for
Fig.~\ref{fig:full}. Insets of both panels show the details of
local order parameters around their SS phases with enlarged
scales. } \label{fig:othersJd}
\end{figure}

As shown in Fig.~\ref{fig:full}, a novel SS phase can exhibit as
ground states of frustrated spin-dimer models when the frustrated
coupling $J_d$ becomes large enough. One may wonder what the fate
of the SS phase is if $J_d$ increases further. In
Fig.~\ref{fig:othersJd}, two values of even larger $J_d$ ($J_d =
0.23$ and 0.27) are considered. We find that the width in $h$ of
the SS state slightly decreases from $h_{c3}-h_{c2} \approx 0.08$
for $J_d = 0.21$ (see Fig.~\ref{fig:enlarged}) down to about 0.06
for $J_d = 0.27$ (See Fig.~\ref{fig:othersJd}). Nevertheless, as
seen from Fig.~\ref{fig:othersJd}, the regime for the plateau
state expands upon increasing $J_d$. Moreover, the width in $h$ of
the SF phase shrinks and the magnitude of $n_0$ in this state
decreases as $J_d$ increases. These outcomes indicate that the SS
state may eventually disappear for large enough frustration. That
is, the SS phase may be stabilized only within a limited region of
the $h$-$J_d$ phase diagram. Therefore, carefully tuning system
parameters into the suggested parameter regime are necessary to
uncover experimentally this novel phase in real frustrated
spin-dimer compounds.


To summarize, the field-induced quantum phase transitions in
frustrated spin-$1/2$ spin-dimer models are studied under the
combined iTEBD and TRG algorithm.~\cite{Jiang08,Chen09} Clear
evidence of a SS phase over a finite regime of magnetic field is
provided. Critical fields at various field-induced transitions are
evaluated accurately by using merely moderate bond dimension $D$.
Besides providing quantitative predictions of the needed
frustrated coupling and magnetic field for the appearance of the
SS phase, the present work also demonstrates clearly the potential
in applying the current formalism even to frustrated spin systems.

We are grateful to K.-K. Ng for simulating discussion.
P. Chen, C.-Y. Lai, and M.-F.Yang thank the support from the NSC of
Taiwan under Contract No. NSC 95-2112-M-007-029-MY3
and NSC 96-2112-M-029-004-MY3 respectively.
This work is supported by NCTS of Taiwan.


\begin{thebibliography}{99}
\bibitem{note}
Around critical fields, we have cross-checked the results by
starting the simulation with different initial inputs to ensure
the obtained quantities describing true ground states, rather than
possible metastable states.
\end{thebibliography}
\end{document}